\documentclass[acus]{JAC2000}

\usepackage[final]{graphics}
\usepackage{citesort}
\newcommand{\GG}        {\ensuremath{\mathrm{\gamma\gamma}}}
\newcommand{\BB}        {\ensuremath{\mathrm{B\overline{B}}}}
\newcommand{\PP}        {\ensuremath{\mathrm{p\overline{p}}}}
\newcommand{\LL}        {\ensuremath{\mathrm{\Lambda\overline{\Lambda}}}}

\newcommand{\Pe}        {\ensuremath{\mathrm{e}}}
\newcommand{\Pap}        {\ensuremath{\mathrm{\overline{p}}}}
\newcommand{\Pp}        {\ensuremath{\mathrm{p}}}
\newcommand{\EE}        {\ensuremath{\mathrm{e^+}\mathrm{e^-}}}
\newcommand{\KK}        {\ensuremath{\mathrm{K^{+}K^{-}}}}

\newcommand{\MuMu}      {\ensuremath{\mathrm{\mu^{+}\mu^{-}}}}
\newcommand{\PiPi}      {\ensuremath{\mathrm{\pi^{+}\pi^{-}}}}
\newcommand{\GV}        {\ensuremath{\mathrm{GeV}}}
\newcommand{\MV}        {\ensuremath{\mathrm{MeV}}}
\newcommand{\costs}     {\ensuremath{\mathrm{|\cos\theta^{*}|}}}
\newcommand{\cost}      {\ensuremath{\mathrm{|\cos\theta|}}}
\newcommand{\wo}        {\ensuremath{w_{1}}}
\newcommand{\wt}        {\ensuremath{w_{2}}}
\newcommand{\wi}        {\ensuremath{w_{i}}}
\newcommand{\X}         {X}
\newcommand{\PT}        {\ensuremath{p_{\perp}}}
\newcommand{\D}         {\ensuremath{\rm d}}
\newcommand{\pb}        {\ensuremath{\rm pb}}
\newcommand{\dedx}      {\ensuremath{{\rm d}E/{\rm d}x}}
\newcommand{\spts}      {\ensuremath{|\sum{\vec{p}_\perp}|^{2}}}
\newcommand{\SI}        {\ensuremath{\mathrm{\sigma}}}

\title{\flushright{W02}\\[15pt] \centering EXPERIMENTAL STATUS OF
PHOTON PHOTON INTO BARYON ANTIBARYON PAIRS\thanks{Work
supported by Department of Energy contract DE-FG03-95ER-40894}}

\author{T. Barillari, Univ. of Colorado, Boulder \\
    {e-mail: \tt Barillari@SLAC.Stanford.EDU}}
\begin{document}

\maketitle

\begin{abstract}
The exclusive production of $\BB$ pairs in the collisions of two 
quasi-real photons have been studied in different experiments at
$\EE$ colliders.
Results are presented for the processes $\GG\to\PP$ and $\GG\to\LL$.
The cross-section measurements are compared      
with the recent analytic calculations based on the quark-diquark
model predictions.  
Monte Carlo studies have been done to investigate the PEP-N
expectations for the $\GG\to\PP$ process.
\end{abstract}

\section{INTRODUCTION}\label{sec:introduction}

The exclusive production of baryon-antibaryon pairs in the collision
of two quasi real photons can be used to test QCD predictions.
The photons are emitted by the beam electrons and positrons and the
$\BB$ are produced in the process $\EE\to\EE\GG\to\EE\BB$.

The application of QCD to exclusive two-photon reactions is based 
on the work of Brodsky and Lapage~\cite{Lepage:1980fj}. According to their 
formalism the process is factorized into a non-perturbative part, 
the hadronic wave function of the final state, and a perturbative 
part. A calculation based on this ansatz~\cite{Farrar:1985gv,Millers:1986ca} 
uses a specific model of the proton's three-quark wave function by Chernyak and 
Zhitnitnitsky~\cite{Chernyak:1984bm}. This calculation predicts cross-sections 
that are about one order of magnitude smaller than the existing 
experimental 
results~\cite{Althoff:1983pf,Bartel:1986sy,Aihara:1987ha,Albrecht:1989hz,Artuso:1994xk,Hamasaki:1997cy,teresa}, for two-photon center-of-mass energies $W$ 
greater than $2.5\,\GV$.

To model non-perturbative effects, the introduction of diquarks has
been proposed~\cite{Ansel:1987vk}. Within this model, baryons are viewed 
as systems of quarks and diquarks, quasi-elementary constituents which 
partially survive medium-hard collision. Their composite nature is taken 
into account by diquark form factors. 
Recent studies~\cite{berger:1997} have extended the investigation of 
exclusive reactions within the diquark model to two-photon 
reactions~\cite{Anselmino:1989gu,Kroll:1991a,Kroll:1993zx,Kroll:1996pv}.

The quark-diquark model works rather well for exclusive reactions in the
space-like region~\cite{Kroll:1991a,Kroll:1991ag,Kroll:1990hg}. 
The calculations of the integrated cross-sections for the processes 
$\GG\to\PP$ and $\GG\to\LL$ in the angular region $\costs < 0.6$, 
$\theta^{*}$ here is the the polar angle of the $\GG$ 
centre-of-mass system (cms), show a good agreement with the existing 
data described in Sec.~\ref{sec:3.0} and Sec.~\ref{sec:4.0} in this paper.
The $\GG\to\PP$ Monte Carlo studies for PEP-N are given 
in Sec.~\ref{sec:5.0}.


\section{THEORY }\label{sec:2.0}

\subsection{Two-photon physics at $\EE$ storage rings}\label{sec:2.1}

The two-photon process is a two step process: first
both incident particles emit virtual photons with squared masses 
$q_1^{2}$, $q_2^{2}$ and energies $\wo$, $\wt$. Next the two 
photons produce the final state $\X$. The first step, the 
${\Pe}{\Pe}\gamma$ vertex, is completely specified by quantum 
electrodynamics (QED); the second step, $\GG\to \X$, is not
rigorously calculable for an hadronic final state $\X$. 
An approximation for the cross-section can be obtained if the 
essential features of both ${\Pe}{\Pe}\gamma$ vertices are given: 
the $1/q_{i}^{2}$-dependence from the photon propagator together with 
the $1/\wi$ dependence characteristic of bremsstrahlung. 

A natural way of differentiating between final states $\X$ produced by the 
two-photon process and those produced by the $\EE$ annihilation 
process is the observation of a scattered electron, called ``tag''.
Depending  on the number of electrons detected (zero, one or two) events 
are referred to as no-tag, single-tag or double-tag, respectively. 
In a no-tag event the scattered electrons go undetected in the beam pipe. 
Consequently, the final-state $\X$ coming from the reaction $\EE\to\EE \X$ 
has a small transverse momentum. 
If $\X$ then decays into two charged particles, usually these particles
are detected at small angles with respect to the beam.
They are back-to-back in the $x-y$ plane, but in the $x-z$ plane they are not. 
The $\GG$ center of mass is moving and boosted along the beam axis. 
The higher the momentum, the closer are the produced particles to the 
beam direction.  
This feature, combined with the typically low mass of two-photon produced 
final states, severely limits the detection efficiency which rarely exceeds 
$10\,\%$.   


\subsection{Hard Scattering Picture (HSP)}\label{sec:2.2}

In the perturbative QCD scheme, also called hard-scattering-picture
(HSP) see 
Refs.~\cite{Brodsky:1975vy,Lepage:1980fj,Brodsky:1981rp,Mueller:1981sg,Chernyak:1984bm,Botts:1989kf}, 
an exclusive hadronic process, $A + B\rightarrow C + D$, to leading order
in the inverse of the large momentum transfer in the transverse direction, 
$1/p_\perp$, is described by an exclusive hadronic amplitude $\mathcal{M}$.
This amplitude  can be expressed as a convolution of 
process-independent distribution amplitudes, $\phi_{H_{i}}$, with the 
elementary scattering amplitude, $T_{H}$
\begin{eqnarray}
\mathcal{M} &=& \int\limits_0^1 T_H(x_j,p_\perp) \prod_{H_i} \left( \phi_{H_i}
(x_j,\tilde{p}_\perp)\phantom{\prod_{j = 1}^{n_i}} \right.\nonumber\\
            & & \left.\times\delta ( 1 - \sum_{k = 1}^{n_i} x_k )\prod_{j = 1}^{n_i} d\,x_j \right),
\label{eq:amplitude}
\end{eqnarray}
where $\tilde{p}_\perp\,\approx min(x_j,1-x_j)\sqrt{s}|\sin\theta|$.

Eq.~(\ref{eq:amplitude}) separates the hard-scattering amplitude from
the bound state dynamics, namely the short-range from the long-range 
phenomena.

One important phenomenological consequence coming from this factorization 
formula is  
the \emph{dimensional counting rules}. 
By ignoring logarithmic corrections~\cite{Brodsky:1973kr,Matveev:1973ra},
the dimensional counting rules predict the following power-law behavior
of the $\GG\to\BB$ (${\rm B}$ = Baryons) cross-section at fixed angles:
\begin{equation}
\left(\frac{\D\sigma_{\GG\to \BB}}{\D t}\right)\sim s^{-6}.
\label{eq:sminuss}
\end{equation}
The scaling law is valid only at sufficiently large $p^{2}_{\perp}$
when $\alpha_{\rm s}(p^{2}_{\perp})$ is small enough to make the Feynman
diagram expansion meaningful.

Another important consequence of Eq.~(\ref{eq:amplitude})  
is the \emph{hadron helicity conservation rules}. 
For each exclusive reaction $A + B\rightarrow C + D$, the sum of the 
initial helicities equals the sum of the final ones~\cite{Brodsky:1981kj}:
\begin{equation}
\lambda_A+\lambda_B=\lambda_C+\lambda_D .
\label{eq:mot8}
\end{equation}

The \emph{dimensional counting rules} are in good 
agreement with the data~\cite{Anderson:1973cc,Stone:1978jh,Arnold:1986nq}. 
However, the \emph{hadron helicity conservation} rule 
has given some troubles when its consequences are compared 
with the existing spin data in exclusive hadronic reactions.
An example of a typical problem that raises from the Eq.~(\ref{eq:mot8}),
comes from the $\eta_{c}$ and $\chi_0$ decays into $\PP$, see 
Refs.~\cite{Baglin:1986br,Balt:1986}.


\subsection{Spin problems and the diquark solution}\label{sec:2.3}

The introduction of diquarks at this point may have two positive
consequences. The first consequence is that it modifies 
the dimensional counting rules of Eq.~(\ref{eq:sminuss}) by effectively 
decreasing the number of constituents to be taken into account in the 
process studied.
The power law behavior of the cross-section for e.g. the $\GG\to\PP$ 
process~\cite{Anselmino:1989gu} is then given by:
\begin{equation}
\left(\frac{\D\sigma_{\GG\to \BB}}{\D t}\right) \sim {s^{-4}}.
\label{eq:ggpp}
\end{equation}

The second consequence of diquarks as constituents has to do with the 
violation of Eq.~(\ref{eq:mot8}). This violation can only  
come from couplings between gluons and those partons that allow helicity 
flips, such as vector diquarks. Again, at very large $Q^2$ values, if a 
diquark resolves into two quarks, the helicity-conservation rule is 
recovered, while at $Q^2$ values where the diquarks act as elementary 
objects helicity conservation can be strongly violated, which solves the 
quark model spin problems.

From all the applications analyzed, see summary given in 
Ref.~\cite{mauro},  it emerges that diquarks seem to be a useful 
phenomenological way of modeling higher-order and non-perturbative 
effects in order to achieve a better description of many hadronic 
exclusive reactions.
Nevertheless, the treatment of exclusive processes in the framework 
of constituent models and perturbative QCD is really far from being 
understood in a unique and well defined computational scheme.


\subsection{Two-Photon annihilation into baryon-anti-baryon pairs}
\label{sec:2.4}

There are recent applications of the quark-diquark model that concern the 
class of reactions $\GG\to {\rm B\overline{B}}$~\cite{berger:1997}, where 
B represents an octet baryon (${\rm B} = \Pp, \Lambda, \Xi$, etc.).
In the older calculations of Ref.~\cite{Anselmino:1989gu} the
$\GG\to\PP$ annihilation has been computed in the scheme of 
Refs.~\cite{Brodsky:1975vy,Lepage:1980fj}. Diquarks
are in this work considered as quasi-elementary constituents, all the masses
are neglected except those of the scalar diquarks in the propagator. 
Within the new calculations of Ref.~\cite{berger:1997}, baryon-mass 
effects are instead taken into account, and the cross-sections have been 
computed down to energy values of $2.2\,\GV$. At these values the diquark model 
starts to loose its validity, but this is where most of the experiments
have their bulk of data.


\section{THE $\GG\to\PP$ PROCESS }
\label{sec:3.0}

There are recent studies for the exclusive $\GG\to\PP$ cross-section 
measurements using the OPAL data at LEP2, $\sqrt{s}=183$ and 
$189\,\GV$, see Ref.~\cite{teresa}.
The $\GG\to\PP$ events are selected in OPAL by applying the 
following main set of cuts:
\begin{enumerate}
\item
     Exactly two oppositely charged tracks; the tracks
     must have at least 20 hits in the central jet chamber. 
     The selected tracks must have a minimal distance, $|d_{0}|$, of at 
     most $1$~cm from the beam axis.
\item
     For each track the polar angle must be in the range $\cost < 0.75$ 
     and the transverse momentum
     $\PT$ must be larger than $400\,\MV$.
     These cuts ensure a high 
     trigger efficiency and good particle identification.
\item
     The polar angle in the $\GG\to\PP$ cms has to be in the range 
     $\costs < 0.6$. 
\item 
     Data and Monte Carlo events must pass a defined trigger 
     condition based on a combination of track and time-of-flight 
     triggers.
\item
     Exclusive two-particle final states are selected by
     rejecting events if the transverse component of the momentum 
     sum squared of the two tracks, $\spts$, is larger than
     $0.1\,\GV^2$.
\item
     The large background from other exclusive processes, 
     mainly the production of $\EE$, $\MuMu$, $\PiPi$ and 
     $\KK$ pairs, is reduced by particle identification using the 
     specific energy loss cuts, $\dedx$. The 
     $\dedx$ probabilities of the tracks must be consistent with the 
     $\Pp$ and $\Pap$ hypothesis. 
\end{enumerate}

Within the applied $\costs < 0.6$ cut, the typical OPAL 
detection efficiency is about $2\%$ at high values of $W$ and
about $7-11\%$ at low $W$. 
Similar values of detection efficiency are also found in other 
experiments: e.g. the TASSO~\cite{Althoff:1983pf} detection
efficiency, was found to be $1.0\pm0.17\%$ at $W=2.0\,\GV$, 
$6.5\pm0.6\%$ at $2.5\,\GV$ and $3.0\pm0.6\%$ at $3.1\,\GV$.

The OPAL trigger efficiency for $\PT>400\,\MV$ and for $W>2.15\,\GV$ 
is about $94\%$.
In VENUS the trigger efficiency for tracks with $\PT\geq 600\,\MV$ 
was about $97\%$.   

\subsection{Cross-section measurements}
\label{sec:3.1}

\begin{figure}[htbp]
\centering
\resizebox{0.4\textwidth}{!}{
\includegraphics{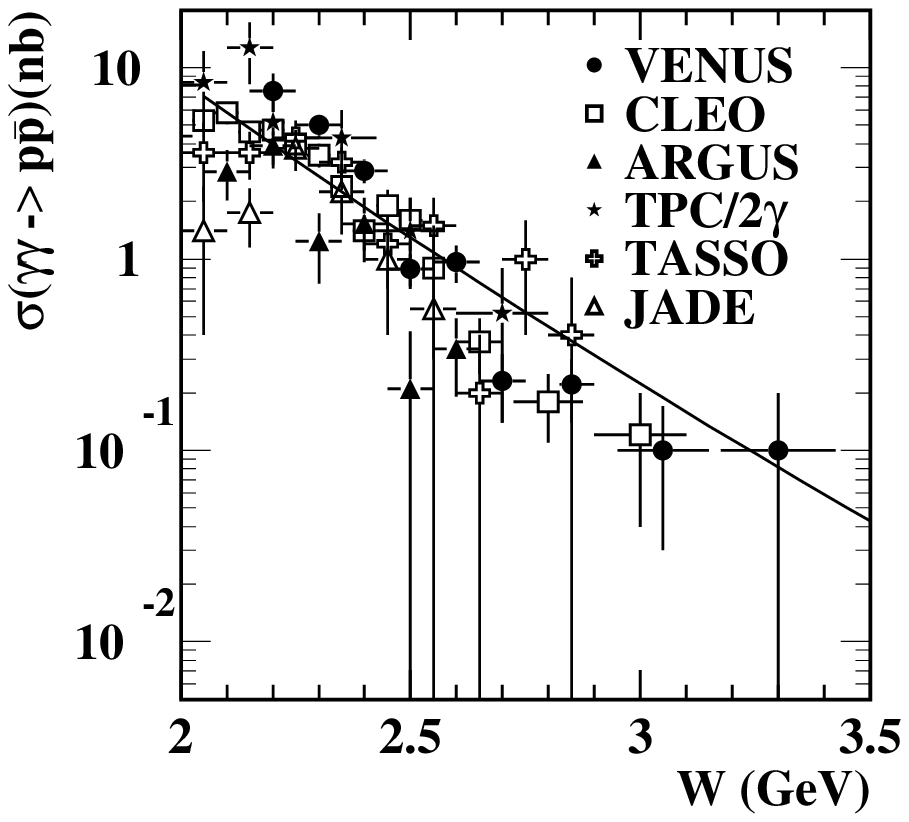}}
\resizebox{0.4\textwidth}{!}{
\includegraphics{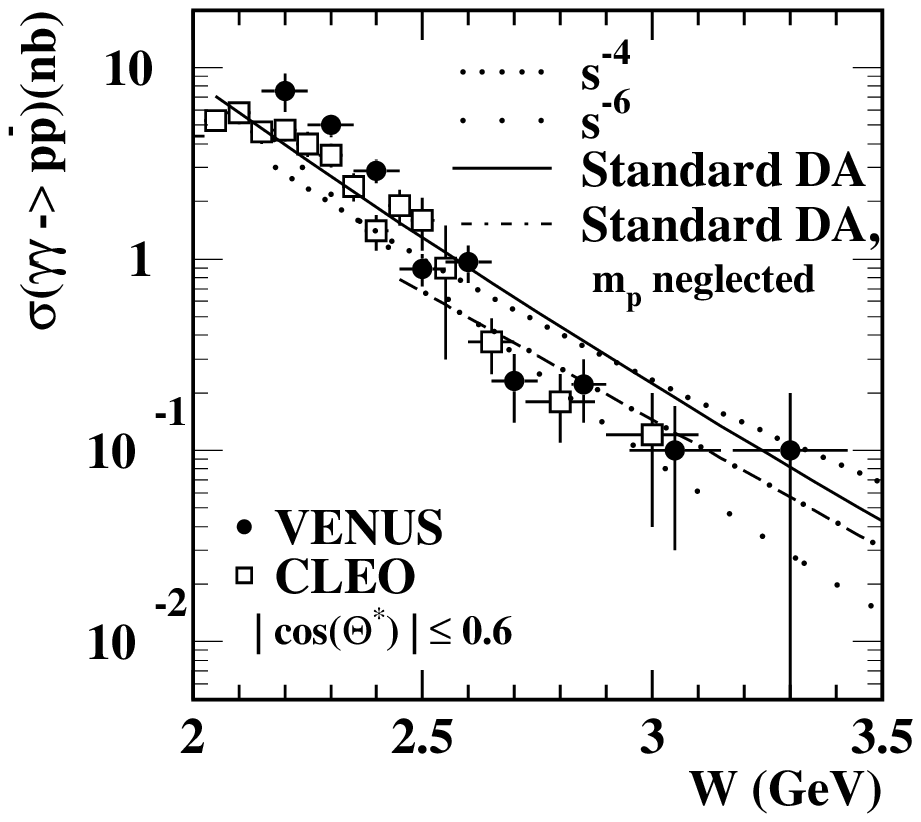}}
\caption{Cross-sections $\sigma(\GG\to\PP)$ as a function of $W$ for
        $\costs < 0.6$. 
        In the figure on top the latest data obtained in 
        VENUS~\cite{Hamasaki:1997cy} and CLEO~\cite{Artuso:1994xk} are compared 
        to other experimental results~\cite{Althoff:1983pf,Bartel:1986sy,Aihara:1987ha,Albrecht:1989hz};      
        and to the quark-diquark model predictions~\cite{berger:1997}
        (solid line).
        Here only statistical errors are shown. In the figure on the bottom 
        the VENUS and CLEO data are shown together with the quark-diquark 
        model of Ref.~\cite{Ansel:1987vk,Kroll:1993zx} (dash-dotted line), 
        and of Ref.~\cite{berger:1997} (solid line), 
        and with the power law predictions with the fixed 
        exponents $-6$, and $-4$. 
        In this picture the error bars are statistical only.}
\label{fig:1}
\end{figure}

The list of the existing exclusive $\GG\to\PP$ cross-section measurements 
are given in Table~\ref{tab:crossmeas}.
\begin{table*}[htb]
 \begin{center}
  \begin{tabular}{l|c c c c c}
   \hline 
    $\EE$     & Year& $E_{Beam}$ &Integrated    & $W_{\GG}$  & Number of\\
    Experiments & &$(\GV)$&Luminosity ($\pb^{-1}$)&  (\GV)& $\PP$ events\\
    \hline    TASSO (DESY) &1982& $15 - 18.3$& $19.685$     & $2.0 - 2.6$& $8$  \\
    TASSO (DESY) &1983& $17$       & $74$         & $2.0 - 3.1$& $72$ \\
    JADE (DESY)  &1986&$17.4 - 21.9$&$59.3 + 24.2$& $2.0 - 2.6$& $41$ \\
    TPC/2$\gamma$ (SLAC) &1987&$14.5$&$75$        &$2.0 - 2.8$&$50$   \\
    ARGUS (DESY) &1989& $4.5 - 5.3$& $234$        &$ 2.6 - 3.0$&$60$  \\
    CLEO (CESR)  &1994& $5.29$     &$1310$        &$2.0 - 3.25$&$484$ \\
    VENUS (TRISTAN) &1997&$57 - 64$&$331$         &$2.2 - 3.3$ &$311$ \\
   \hline
   \hline
    PEP-N (SLAC) & - & $0.5_{(VLER)}-3.1_{(LER)}$& $200$     & $1.9 - 2.6$& $60$  \\
   \hline
  \end{tabular}
  \caption{ The experiments that have measured the
      $\GG\to\PP$ cross section in $\EE$ collision;
      the table gives the beam energy, the
      integrated luminosity, the range of $W$
      and the total number of $\PP$ events.
      The last line summarizes the PEP-N expectation.}
  \label{tab:crossmeas}
  \end{center}
\end{table*}
The OPAL data are not published yet, therefore they are not shown
here.
In Fig.~\ref{fig:1} (top) the latest VENUS~\cite{Hamasaki:1997cy} 
and CLEO~\cite{Artuso:1994xk} results
are compared to the cross-section measurements obtained by 
TASSO~\cite{Althoff:1983pf}, JADE~\cite{Bartel:1986sy},
TPC/$2\gamma$~\cite{Aihara:1987ha}, ARGUS~\cite{Albrecht:1989hz}. 
Also shown in the figure are the quark-diquark model 
predictions~\cite{berger:1997}.
A large spread of data is visible in this figure. The 
CLEO~\cite{Artuso:1994xk} results can be considered the
most precise measurements, and they lie in the center of the other data
points.


Fig.~\ref{fig:1} (bottom) shows the VENUS and 
CLEO cross-section measurements as function of $W$ together with the 
most recent quark-diquark model predictions~\cite{berger:1997} 
(solid line in the figure) and the previous calculations of 
Refs.~\cite{Ansel:1987vk,Kroll:1993zx} (dash-dotted line).  

This figure shows that at low $W$ the VENUS measurements are larger 
than the CLEO results.
There is good agreement between these two measurements in the 
higher $W$ region: $W>2.6\,\GV$.
The CLEO results show here a good agreement with the most recent 
quark-diquark model~\cite{berger:1997} in the low invariant mass region, 
the VENUS results lie instead above these predictions.
In the higher $W$ region the VENUS and CLEO data lie below the 
predictions of Ref.~\cite{berger:1997}.  
Within the statistical errors these measurements, 
in the high invariant mass region, can be considered in 
agreement with the calculations of Refs.~\cite{Ansel:1987vk,Kroll:1993zx}. 
The preference of either quark-diquark model of Ref.~\cite{berger:1997} and of 
Refs.~\cite{Ansel:1987vk,Kroll:1993zx} respectively is not obvious  
from the results shown in this figure. 

Finally, the power law predictions of Eq.~(\ref{eq:sminuss}), for $W^2 = s$ 
using the fixed exponents $-6$, and $-4$ are also shown in 
Fig.~\ref{fig:1} (bottom). 
More data at both higher and lower values of $W$ are needed to determine 
which is the correct power law to choose to describe the data. 

\begin{figure}[htbp]
\centering
\resizebox{0.4\textwidth}{!}{
\includegraphics{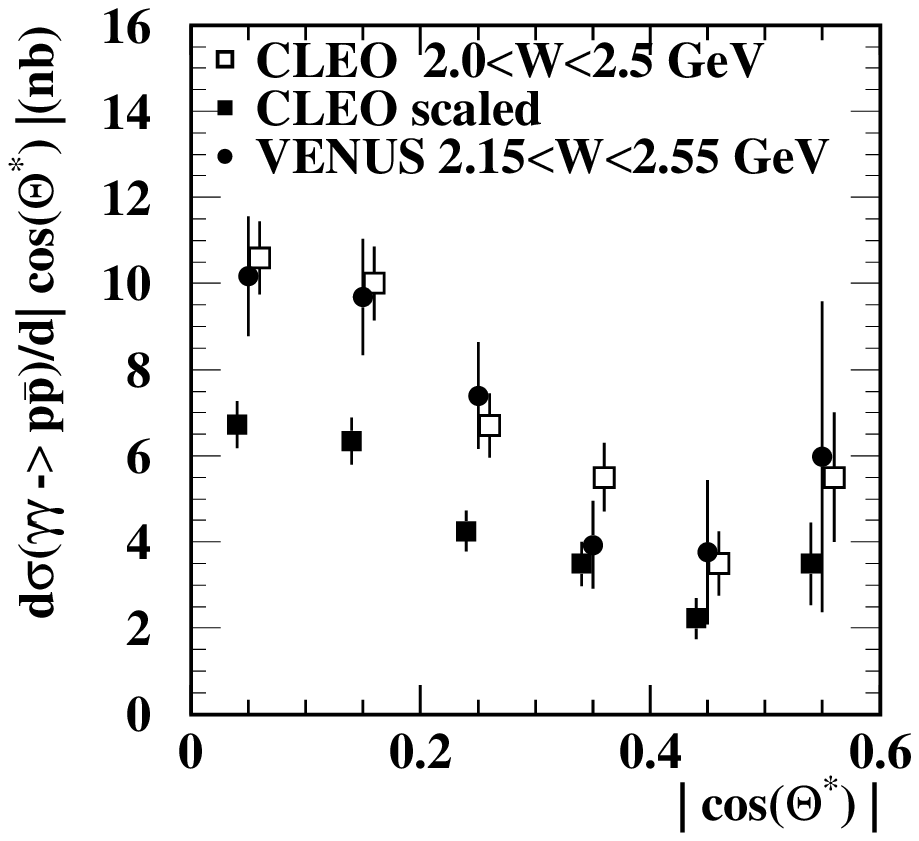}}
\resizebox{0.4\textwidth}{!}{
\includegraphics{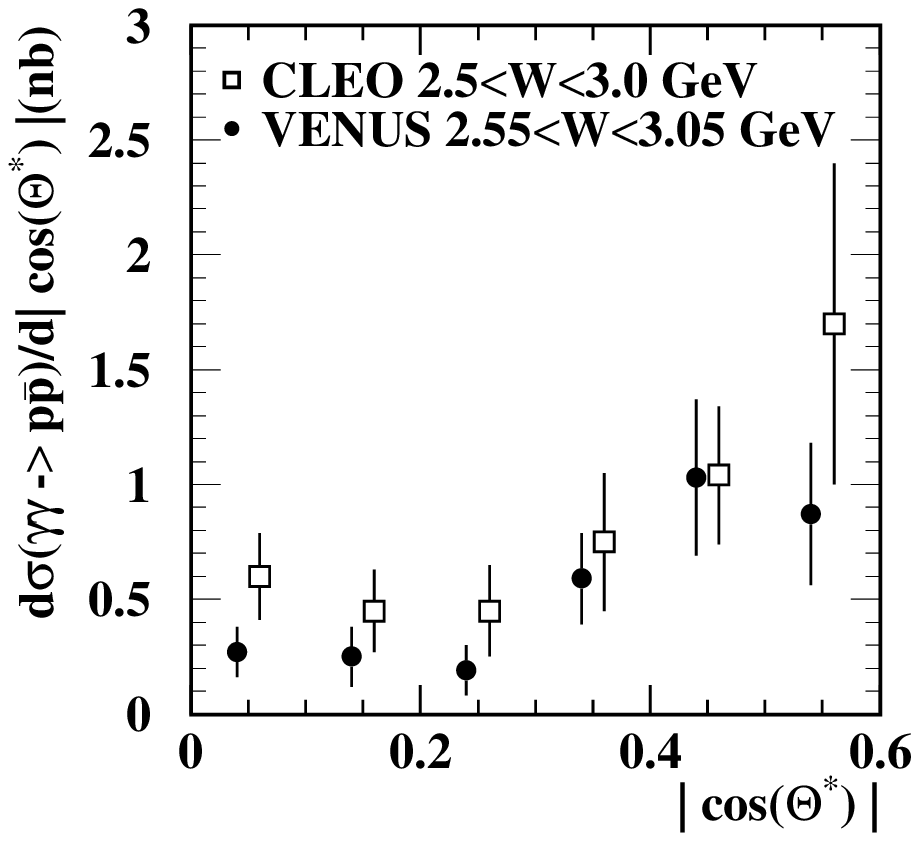}}
\caption{Differential cross-sections for $\GG\to\PP$ as a function of 
        $\costs$. Data from Refs.~\cite{Artuso:1994xk,Hamasaki:1997cy}
        are for a low range of $W$, $2.15\,\GV<W<2.55\,\GV$ (top), 
        and the high range of $W$ (bottom), $2.5\,\GV<W<3.0\,\GV$
        for CLEO and $2.55\,\GV<W<3.05\,\GV$ for VENUS.
        Errors are statistical only.}
\label{fig:2}
\end{figure}

Fig.~\ref{fig:2} (top) shows the VENUS and CLEO measured differential 
cross-sections as function of $\costs$ in the range of 
$2.15\,\GV<W<2.55\,\GV$. 
In the two experiments the differential cross-section decreases 
toward $\costs=0.6$. The scaled CLEO measurement lies below the VENUS 
results.
The scaling factor used to shift the CLEO differential cross-section 
measurements from the range of $W$ between $2.0 - 2.5\,\GV$ to 
the range of $W  = 2.15 - 2.55\,\GV$ used by VENUS, is $0.6345$.  
This scaling factor is computed by dividing the two CLEO 
total cross-sections integrated over the two considered 
$W$ ranges of $2.0 - 2.5\,\GV$ and $2.15 - 2.55\,\GV$.

Fig.~\ref{fig:2} (bottom) shows the measured differential cross-section 
as function of $\costs$ in the high $W$ region:
$2.55\,\GV<W<3.05\,\GV$ for VENUS and $2.5\,\GV<3.0\,\GV$ for CLEO. 
There is good agreement between the results obtained by these two
experiments.

\begin{figure}[htbp]
\centering
\resizebox{0.4\textwidth}{!}{
\includegraphics{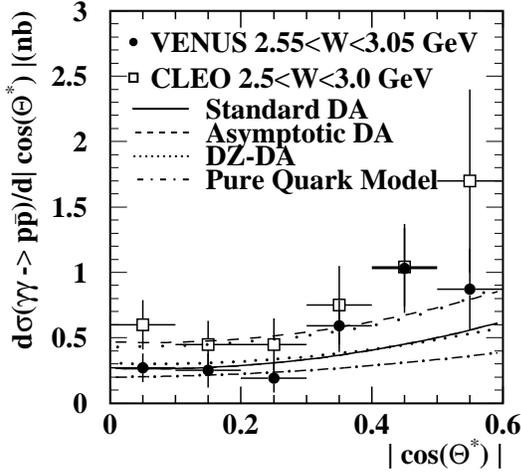}}
\caption{VENUS and CLEO differential cross-section 
      ${\rm d}\SI{(\GG\to\PP)}/{\rm d}\costs$, shown with statistical
      and systematic errors, 
      in the range $2.55\,\GV <W< 3.05\,\GV$ for VENUS and
      $2.5\,\GV <W< 3.0\,\GV$ for CLEO compared to
      the theoretical predictions given in Ref.~\cite{Farrar:1985gv} 
      (dash-dotted line), in Ref.~\cite{Ansel:1987vk,Kroll:1993zx} 
      (dotted line), and in Ref.~\cite{berger:1997} (the other lines) 
      for $\costs < 0.6$.}
\label{fig:3}
\end{figure}

Fig.~\ref{fig:3} (top) shows the comparison of the VENUS and CLEO  
differential cross-section measurements for the higher $W$ region 
with the calculation given in Ref.~\cite{berger:1997} at $W = 2.8\,\GV$ 
for different distribution amplitudes (DA).
The results of the pure quark model~\cite{Farrar:1985gv,Millers:1986ca} 
are also shown.
The pure quark model and the quark-diquark model predictions, 
lie below the data but in both cases the shape is reasonably 
well reproduced. This could indicate that the
\emph{hadron helicity conservation rules} of Eq.~(\ref{eq:mot8}) are
satisfied in the high $W$ region but they are not in the low 
$W$ values. In fact, in this low $W$ region the measured differential 
cross-sections have a different distribution from the differential 
cross-sections obtained in the high invariant mass region.

The different shape of the curves in these figures shows also 
that for low $W$ the perturbative calculations
of~\cite{Farrar:1985gv,Millers:1986ca} are not valid and 
the $\PP$ system might be described as a bound system with orbital 
angular momentum greater than zero.


\section{The $\GG\to\LL$ Process}
\label{sec:4.0}

\begin{figure}[htbp]
\centering
\resizebox{0.55\textwidth}{!}{
\includegraphics{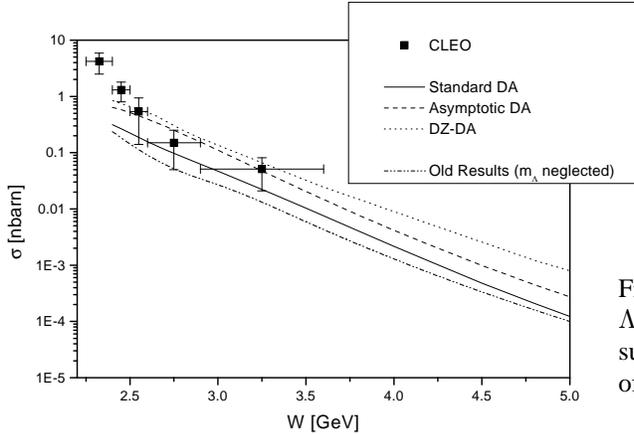}}
\caption{CLEO~\cite{Anderson:1997ak} integrated cross-section for $\GG\to\LL$
         measurement compared with the theoretical results of 
         Ref.~\cite{berger:1997,Kroll:1991a}}
\label{fig:4}
\end{figure}

\begin{figure}[htbp]
\centering
\resizebox{0.4\textwidth}{!}{
\includegraphics{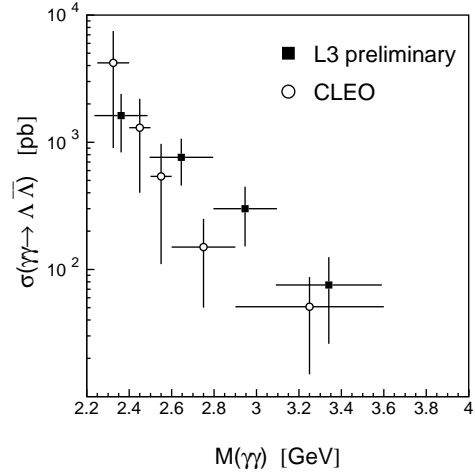}}
\resizebox{0.4\textwidth}{!}{
\includegraphics{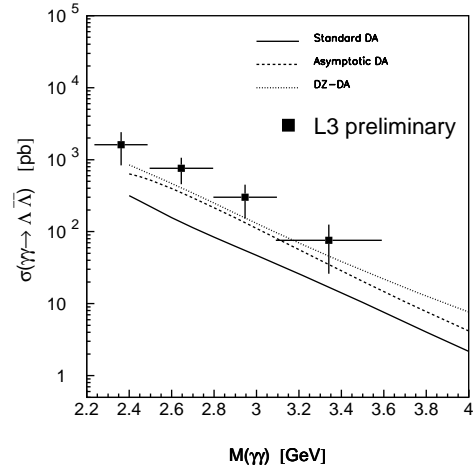}}
\caption{The L3~\cite{l3} integrated cross-section
         $\sigma(\GG\to\LL\X)$ is compared with 
         the CLEO $\sigma(\GG\to\LL$ measurements (top) and
         the quark-diquark model predictions 
         of Ref.~\cite{berger:1997} (bottom).}
\label{fig:5}
\end{figure}

The exclusive cross-section measurement for the $\GG\to\LL$ process and the
inclusive reaction $\EE\to\EE\LL\X$ have been studied by
CLEO~\cite{Anderson:1997ak} and by L3~\cite{l3}, respectively.
The results of the integrated $\GG\to\LL$ cross-section ($\costs<0.6$) 
obtained by CLEO~\cite{Anderson:1997ak} (Fig.~\ref{fig:4})
show a better agreement with the most recent quark-diquark 
predictions~\cite{berger:1997} than compared with the old results of 
Refs.~\cite{Kroll:1991a}. In the low invariant masses region
the data shows a discrepancy with the model. This discrepancy
can be explained by the lower limit of applicability of the 
quark-diquark model itself~\cite{berger:1997}. 
In Fig.~\ref{fig:5} (top) the L3 cross-section 
measurements~\cite{l3} $\SI(\GG\to\LL\X)$ are shown together with 
the CLEO results.
The two measurements can be considered in agreement within large errors.
The comparison of the L3~\cite{l3} data with the most recent quark-diquark 
model predictions~\cite{berger:1997} for the three different distribution
amplitudes is shown in Fig.~\ref{fig:5} (bottom).
The L3 measurements lie above but still in agreement with the predictions.
The excess shown in the data may be due to the $\Sigma^{0}$ 
$\overline{{\Sigma}^{0}}$ and other baryons contamination not removed
from the sample of events analyzed. 

\section{PEP-N expectations}
\label{sec:5.0}

\begin{figure}[htbp]
\centering
\resizebox{0.4\textwidth}{!}{
\includegraphics{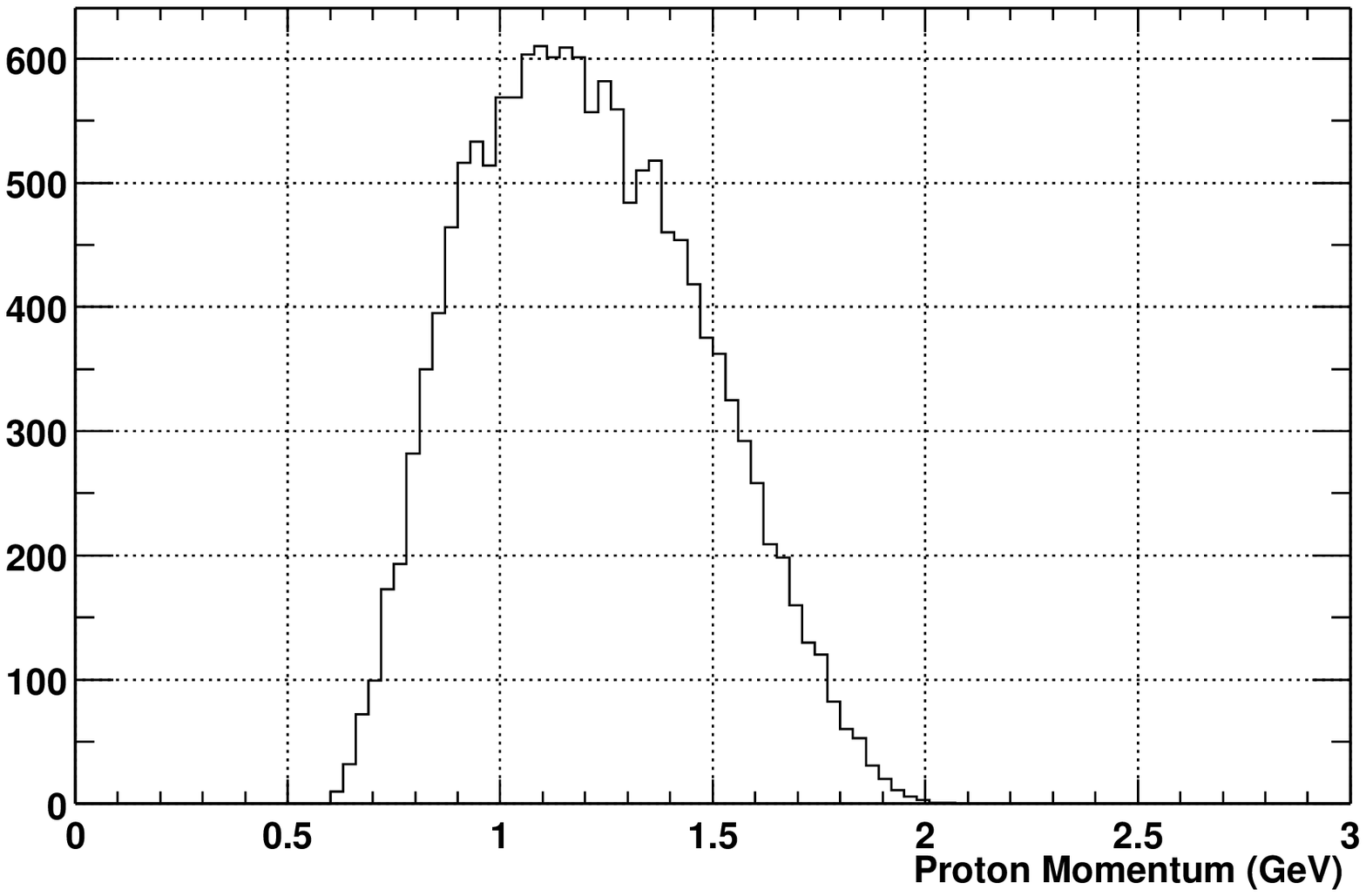}}
\resizebox{0.4\textwidth}{!}{
\includegraphics{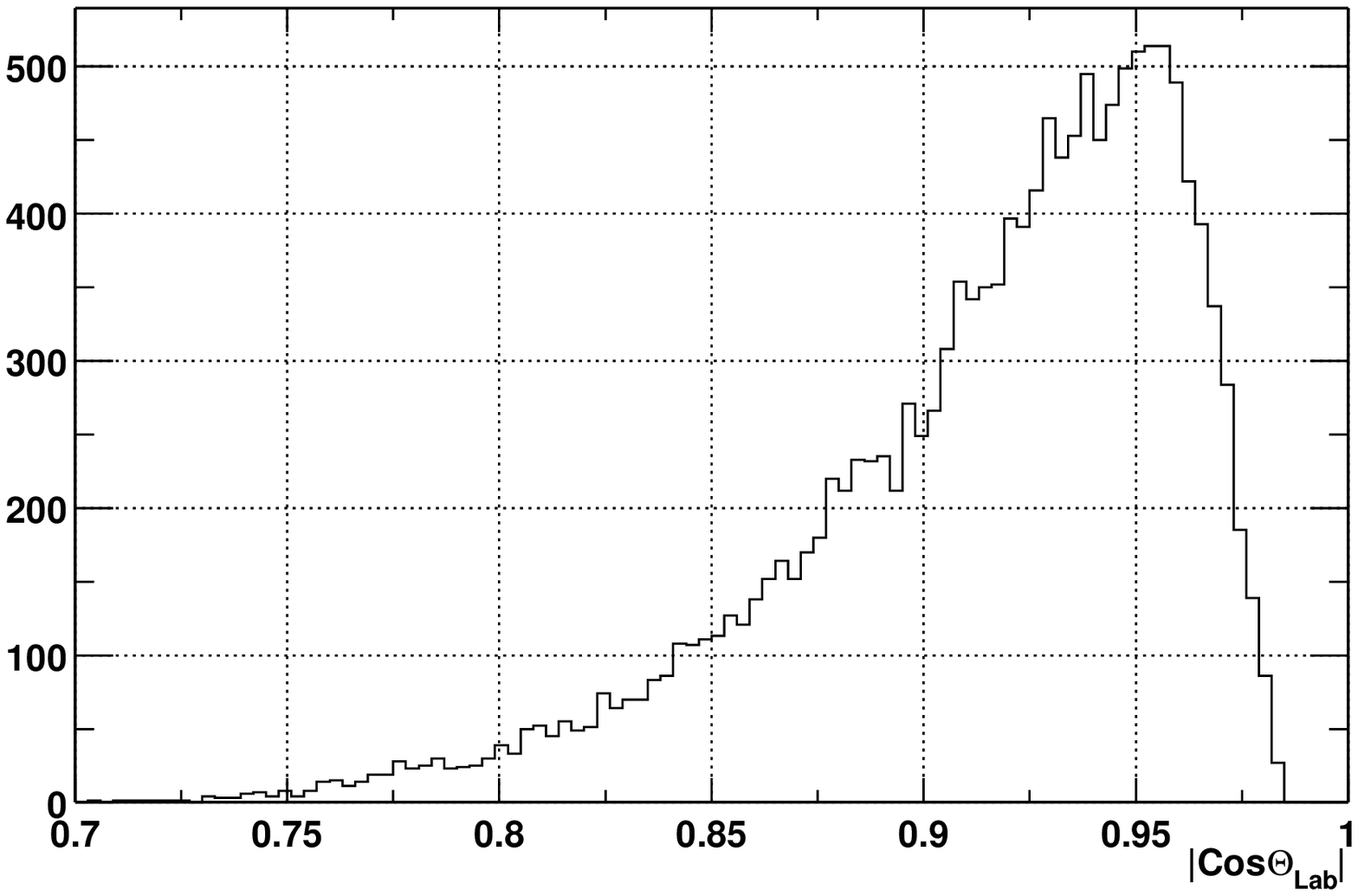}}
\caption{Monte Carlo events distributions for PEP-N at 
        $\sqrt{s} = 2.5\,\GV$ and for beam energies of 
        $3.1\,\GV_{LER}$, $0.5\,\GV_{VLER}$ : 
         (top) proton momentum distribution;  
         bottom $|\cos\theta_{LAB}|$ distribution.}
\label{fig:pepn}
\end{figure}

To understand the possibility of selecting two-photon events and in
particular $\GG\to\PP$ events at PEP-N, preliminary Monte Carlo 
distributions have been studied.
Some quantities are plotted in Fig.~\ref{fig:pepn}.
The $\GG\to\PP$ Monte Carlo events have been simulated with the 
{\sc Galuga}~\cite{Schuler:1996gt,Schuler:1997ex} generator 
within a range of $W$ between $2$ and $2.5\,\GV$.
Due to the beam asymmetry the $\GG$ cms receives a larger boost compared
to a symmetric $\EE$ machine and therefore the
momenta of the final state particles are larger.
Fig.~\ref{fig:pepn} (top) shows that the proton momentum distribution
varies between $0.6-2.0\,\GV$ instead e.g. of the range $0.4-1.1\,\GV$ 
observed for the proton momenta in OPAL~\cite{teresa}.
Fig.~\ref{fig:pepn} (bottom) shows the $|\cos\theta_{LAB}|$~\footnote{$\theta_{LAB}$ 
is the polar angle in the laboratory} distribution.
These two distributions show the better experimental conditions
expected at PEP-N for two-photon events. 
A high detection efficiency, large angular acceptance, 
and a good trigger efficiency due to the higher momentum tracks are anticipated.
The last row in Table~\ref{tab:crossmeas} gives the number of 
$\GG\to\PP$ events expected to be detected at PEP-N under the assumption 
of a good trigger and detection efficiency and for a 
total integrated luminosity of $200\,\pb^{-1}$.

\section{Conclusion}
\label{sec:6.0}

The data shown in this paper indicate that there is still a lot to investigate 
about the exclusive $\GG\to\BB$ processes.
The expected good experimental conditions at PEP-N would make it  
the ideal place to continue these studies, especially in the low 
invariant mass region. 

\section{Acknowledgment}
\label{sec:7.0}

I would like to thank R.~Baldini and S.J.~Brodsky for gently inviting me 
to present this work at the PEP-N workshop.


\begin{thebibliography}{9}


\bibitem{Lepage:1980fj}
S.~J. Brodsky and G.~P. Lepage ,
\newblock Phys. Rev. {\bf D22} (1980) 2157.

\bibitem{Farrar:1985gv}
G.~R. Farrar, E.~Maina, and F.~Neri,
\newblock Nucl. Phys. {\bf B259} (1985) 702.

\bibitem{Millers:1986ca}
J.~F. Gunion and D.~Millers,
\newblock Phys. Rev. {\bf D34} (1986) 2657.

\bibitem{Chernyak:1984bm}
V.~L. Chernyak and I.~R. Zhitnitsky,
\newblock Nucl. Phys. {\bf B246} (1984) 52.

\bibitem{Althoff:1983pf}
TASSO, M.~Althoff and others,
\newblock Phys. Lett. {\bf B130} (1983) 449.

\bibitem{Bartel:1986sy}
JADE, W.~Bartel and others,
\newblock Phys. Lett. {\bf B174} (1986) 350.

\bibitem{Aihara:1987ha}
TPC/Two Gamma, H.~Aihara and others,
\newblock Phys. Rev. {\bf D36} (1987) 3506.

\bibitem{Albrecht:1989hz}
ARGUS, H.~Albrecht and others,
\newblock Z. Phys. {\bf C42} (1989) 543.

\bibitem{Artuso:1994xk}
CLEO, M.~Artuso and others,
\newblock Phys. Rev. {\bf D50} (1994) 5484.

\bibitem{Hamasaki:1997cy}
VENUS, H.~Hamasaki and others,
\newblock Phys. Lett. {\bf B407} (1997) 185.

\bibitem{teresa}
T.~Barillari,
\newblock Title: Cross-Section Measurements of the Process $\GG\to\PP$
in Untagged Events at $\sqrt{s} = 183$ and $189\,\GV$ with the OPAL
Detector at LEP, PhD thesis N$^{\circ}$ 3256, University of Geneva (2001).

\bibitem{Ansel:1987vk}
M.~Anselmino, P.~Kroll, and B.~Pire,
\newblock Z. Phys. {\bf C36} (1987) 89.

\bibitem{berger:1997}
C.~F. Berger,
\newblock Title: Exclusive two-photon reactions in the few{\rm GeV} region,
  Diploma thesis, Technological University Graz  (1997).

\bibitem{Anselmino:1989gu}
M.~Anselmino, F.~Caruso, P.~Kroll, and W.~Schweiger,
\newblock Int. J. Mod. Phys. {\bf A4} (1989) 5213.

\bibitem{Kroll:1991a}
P.~Kroll, M.~Schurmann, and W.~Schweiger,
\newblock Int. J. Mod. Phys. {\bf A6} (1991a) 4107.

\bibitem{Kroll:1993zx}
P.~Kroll, Th.~Pilsner, M.~Sch\"{u}rmann, W.~Schweiger, 
\newblock Phys.~Lett. {\bf B316} (1993) 546

\bibitem{Kroll:1996pv}
P.~Kroll, M.~Sch\"{u}rmann, and P.~A.~M. Guichon,
\newblock Nucl. Phys. {\bf A598} (1996) 435.

\bibitem{Kroll:1991ag}
P.~Kroll, M.~Schurmann, and W.~Schweiger,
\newblock Z. Phys. {\bf A342} (1992) 429.

\bibitem{Kroll:1990hg}
P.~Kroll, M.~Schurmann, and W.~Schweiger,
\newblock Z. Phys. {\bf A338} (1991) 339.

\bibitem{Brodsky:1975vy}
S. J. Brodsky and G. R. Farrar,
\newblock Phys. Rev. {\bf D11} (1975) 1309.

\bibitem{Brodsky:1981rp}
S. J. Brodsky and G. P. Lepage,
\newblock Phys. Rev. {\bf D24} (1981) 1808.

\bibitem{Mueller:1981sg}
A. H. Mueller,
\newblock Phys. Rept. {\bf 73} (1981) 237.

\bibitem{Botts:1989kf}
J. Botts and G. Sterman,
\newblock Nucl. Phys. {\bf B325} (1989) 62.

\bibitem{Brodsky:1973kr}
S.~J. Brodsky and G.~R. Farrar,
\newblock Phys. Rev. Lett. {\bf 31} `(1973) 1153.

\bibitem{Matveev:1973ra}
V.~A. Matveev, R.~M. Muradian, and A.~N. Tavkhelidze,
\newblock Nuovo Cim. Lett. {\bf 7} (1973) 719.

\bibitem{Brodsky:1981kj}
S.~J. Brodsky and G.~P. Lepage,
\newblock Phys. Rev. {\bf D24} (1981) 2848.

\bibitem{mauro}
M.~Anselmino and others,
\newblock Rev. Mod. Phys. {\bf 65 No. 4} (1993) 

\bibitem{Anderson:1973cc}
R. L. Anderson and others,
\newblock Phys. Rev. Lett. {\bf 30} (1973) 627.

\bibitem{Stone:1978jh}
J.~L.~Stone and J.~P.~Chanowski and H.~R.~Gustafson and 
M.~J.~Longo and S.~W.~Gray
\newblock Nucl. Phys. {\bf B143} (1978) 1.

\bibitem{Arnold:1986nq}
R. G. Arnold and others,
\newblock Phys. Rev. Lett. {\bf 57} (1986) 174.

\bibitem{Baglin:1986br}
C. Baglin and others,
\newblock Phys. Lett. {\bf B172} (1986) 455.

\bibitem{Brodsky:1987nt}
S.~J.~Brodsky and F.~C.~Erne and P.~H.~Damgaard and P.~M.~Zerwas
\newblock Conribution to ECFA Workshop LEP200, Aachen, Germany
Sep 29 - Oct 1, (1986).

\bibitem{Balt:1986}
R. M. Baltrusaitis and others,
\newblock Phys. Rev. {\bf D33} (1986) 629.

\bibitem{Budnev:1974de}
V.~M. Budnev, I.~F. Ginzburg, G.~V. Meledin, and V.~G. Serbo,
\newblock Phys. Rep. {\bf 15} (1974) 181.

\bibitem{Schuler:1996gt}
G.~A.~Schuler,
\newblock hep-ph/9611249 (1996).

\bibitem{Schuler:1997ex}
G.~A.~Schuler,
\newblock Comput. Phys. Commun. {\bf 108} (1998) 279.

\bibitem{Anderson:1997ak}
CLEO, S.~Anderson and others,
\newblock Phys. Rev. {\bf D56} (1997) 2485.

\bibitem{l3}
L3, M.~Acciarri and others,
\newblock L3 Note {\bf 2566} (2000) 

\end{thebibliography}
\end{document}